\documentclass[5p,number]{elsarticle}
\usepackage{epsfig}
\usepackage{subfigure}
\usepackage{amsmath}
\usepackage{color}
\usepackage{verbatim}
\usepackage{amssymb}
\usepackage{setspace}
\usepackage{graphicx}
\usepackage{subfigure}
\usepackage{dcolumn}
\usepackage{bm}
\usepackage{times}
\usepackage{enumerate}
\input epsf

\graphicspath{{figures/}}

\begin{document}

\author[CU]{Per Sebastian Skardal}
\ead{skardal@colorado.edu} 
\author[CU]{Dane Taylor}
\ead{dane.taylor@colorado.edu}
\author[CU]{Juan G. Restrepo} 
\ead{juanga@colorado.edu}
\address[CU]{Department of Applied Mathematics, University of Colorado at Boulder, Colorado 80309, USA}

\title{Complex macroscopic behavior in systems of phase oscillators with adaptive coupling}


\begin{abstract}
Using recent dimensionality reduction techniques in large systems of coupled phase oscillators exhibiting bistability, we analyze complex macroscopic behavior arising when the coupling between oscillators is allowed to evolve slowly as a function of either macroscopic or local system properties. For example, we observe macroscopic excitability and intermittent synchrony in a system of time-delayed Kuramoto oscillators with Hebbian and anti-Hebbian learning. We demonstrate the robustness of our findings by considering systems with increasing complexity, including time-delayed oscillators with adaptive network structure and community interaction, as well as a system with bimodally distributed frequencies.

\end{abstract}


\maketitle

\section{Introduction}

Large systems of coupled oscillators are ubiquitous in nature and serve as a basis to study collective behavior. Some examples include synchronized flashing of fireflies \cite{FireFly}, cardiac pacemaker cells \cite{Pacemaker}, walker-induced oscillations of the Millennium Bridge \cite{Millenium}, Josephson junction circuits \cite{Marvel1}, audiences clapping \cite{clapping}, circadian rhythms in mammals \cite{Circadian}, cell function \cite{elana}, neural processing \cite{Deco1}, and chemical oscillations \cite{Kuramoto1,Kiss1}. In certain situations these oscillators can be approximately described in terms of only their phase angle $\theta$. Kuramoto showed \cite{Kuramoto1} that the evolution of the phases in an ensemble of $N$ weakly coupled oscillators approximately obeys
\begin{equation}
\dot{\theta}_n=\omega_n+\sum_{m=1}^N H_{nm}(\theta_m-\theta_n),
\end{equation}
where $\theta_n$ and $\omega_n$ are, respectively, the phase and intrinsic frequency of oscillator $n$, and $H_{nm}$ is a $2\pi$-periodic function that describes the coupling between oscillators $n$ and $m$. When such oscillators represent limit cycles arising from a Hopf bifurcation their coupling is generically sinusoidal, leading to the choice $H_{nm}(\theta)=(k_{nm}/N)\sin(\theta)$. When coupling is uniform, i.e. $k_{nm}=k$, one obtains the classical Kuramoto model which has become a paradigm for the study of synchrony in coupled heterogeneous oscillators. 
Generalizations of the Kuramoto model have become an important area of recent research, including investigations of non-sinusoidal coupling \cite{Daido1}, cluster synchrony \cite{seb}, the effects of network topology \cite{JGR1,Pikovsky1}, non-local coupling \cite{Martens1}, external forcing \cite{Childs1}, coupled excitable oscillators \cite{Alonso1}, phase resetting \cite{Levajic1}, time-dependent connectivity \cite{So1}, noise \cite{Nagai1}, and communities of coupled oscillators \cite{seb2,Kawamura1}. Recently, the analysis of many such systems has been simplified by a dimensionality reduction proposed by Ott and Antonsen \cite{OA1,OA2}, making many cases analytically tractable for the first time. Other recent work on dimensionality-reduction methods for large ensembles of phase oscillators includes Refs.~\cite{Marvel2009Chaos,Rosenblum2011PhysicaD}.

A major difficulty in the study of complex systems (e.g., neural processing and cell function) is overcoming the common disconnect between simple microscopic and complex macroscopic dynamics referred to as emergence. In this paper, we study emergent macroscopic behavior that cannot be deduced from the individual oscillator dynamics alone, but requires a systems-level analysis.
We study macroscopic dynamics arising when slow coupling adaptation is combined with large systems of oscillators. 
%
%
Natural examples of systems involving adaption of system parameters such as coupling strength include clapping audiences \cite{clapping}, brain fluctuations \cite{Deco1}, 
regulation of sleep and circadian rhythms \cite{Saper1}, and regulation of cardiac behavior \cite{Rosati1}. One natural way to model adaptive dynamics in such systems is to allow for the connectivity to evolve as a function of the degree of synchrony of the system. 
Recent studies on adaptive oscillator systems have largely modeled two types of synaptic plasticity: spike-timing-dependent plasticity \cite{STDP} and Hebbian learning \cite{Hebbian}.
%
We further classify such adaptation rules as either {\it uniform adaptation} (the evolution of global coupling in all-to-all coupled systems according to global system properties) or {\it network adaptation} (the evolution of individual links in possibly heterogeneous networks according to local properties), both of which are studied here.

The inclusion of adaptive rules in Kuramoto-type systems can result in rich dynamics that has sometimes been proposed to model information processing in the brain~\cite{STDP,Hebbian}. Typically, however, these adaptive rules are added to the standard Kuramoto model, which has relatively simple macroscopic dynamics (e.g. no memory). In this paper, we explore the addition of adaptive rules to oscillator systems exhibiting bistability, such as those studied in Refs.~\cite{Lee1,Martens2,dane,inertia}. When combined with adaptation, we find that bistability allows for complex macroscopic behavior such as excitable and intermittently synchronous states in addition to simple steady-state behavior. In this paper we consider the case where the timescale of coupling adaptation is much larger than the timescale of oscillator dynamics, which will allow us to separate time scales, first solving for the fast oscillator dynamics using the work of Ott and Antonsen \cite{OA1} and then analyzing the slow adaptation dynamics. We find that even when the adaptation is chosen to be a simple function of the system state, a variety of macroscopic behaviors can be attained by varying parameters of that simple function. The dynamics described in this paper fall within the framework of \emph{dynamic bifurcation} theory~\cite{DynBif}, which describes bifurcations that occur in fast dynamics in response to one or more slowly-changing parameters. In this paper the bifurcations correspond to transitions between macroscopic incoherent and synchronized states in response to one or more slowly changing coupling strengths.


This paper is organized as follows. In Sec.~2 we study a system of time-delayed oscillators subject to uniform adaptation. In Sec.~3 we study three more complicated models that yield richer dynamics: (A) network adaptation on a system of time-delayed oscillators, (B) community interaction between two communities of time-delayed oscillators with community-wise uniform adaptation, and (C) uniform adaptation on a system of oscillators with intrinsic frequencies drawn from a bimodal distribution. Finally, in Sec.~4 we conclude by discussing our results.

\section{Time-delayed oscillators with uniform adaptation}

In this section we study a system of $N$ oscillators coupled through a time-delayed order parameter in which the coupling strength is allowed to slowly adapt in response to the values of this order parameter. This system allows for analytic results that will later serve as a guide to the analysis of more complicated systems. Letting $\omega_n$ denote the intrinsic frequency of oscillator $n$ [which we assume to be randomly drawn from a distribution $g(\omega)$] and $r = \frac{1}{N}\sum_{n=1}^N e^{i \theta_n}$ denote the Kuramoto order parameter, we consider the following model,
\begin{align}
\dot{\theta}_n &=\omega_n+k \mbox{Im}(z e^{-i\theta_n}),\label{thetadot}\\
\tau\dot{z} &=r-z.\label{delayed}
\end{align} 
Since Eq.~(\ref{delayed}) can be written as $z(t) = \tau^{-1}\int_{-\infty}^tr(t')e^{\frac{t'-t}{\tau}}dt'$, $z$ may be interpreted as a time-delayed version of $r$. This time-delayed order parameter is the one that affects individual oscillators. In the continuum limit, $N\to\infty$, this system has been shown to represent exactly the case where the coupling between pairs of oscillators in the Kuramoto model is time-delayed with time delays that have an exponential distribution with average $\tau$ \cite{Lee1}. Note that $\tau\to0$ yields $z=r$, which recovers the Kuramoto model \cite{Kuramoto1}. We extend this system by allowing the uniform coupling constant $k$ to slowly adapt following 
\begin{align}
T\dot{k} &=G(k,z)\label{kdot},
\end{align} 
where $T$ is the timescale of adaptation and $G$ is a function that describes the adaptation of $k$ in terms of its current value and the perceived (delayed) order parameter. We will assume that $T$ is much larger than both $\tau$ and the time scale of oscillator dynamics, given by the inverse of the spread of $g(\omega)$ \cite{OA2} so that we may utilize a separation of time scales to solve Eqs.~(\ref{thetadot}) and (\ref{delayed}) assuming constant $k$ and then solve Eq. (\ref{kdot}) while assuming a steady state. Note that letting $T\to\infty$ recovers the non-adaptive system with fixed $k$.

\subsection{Fast oscillators in the continuum limit}

We begin by describing the steady-state collective dynamics of Eqs.~(\ref{thetadot}) and (\ref{delayed}) for fixed $k$ in the continuum limit.
%
%
%
%
%
%
%
We let $f(\omega, \theta,t)$ denote the density of oscillators with frequency $\omega$ and phase $\theta$ at time $t$. Conservation of oscillators implies that $f(\omega, \theta,t)$ must satisfy the continuity equation 
\begin{equation}
\partial_t f +\partial_{\theta}(f \dot \theta)  = 0 . \label{continuity}
\end{equation}
Following Ref.~\cite{Lee1}, this partial differential equation (PDE) can be reduced to the single complex-valued ordinary differential equation (ODE)
\begin{align}
&\dot{r}+(\Delta-i\omega_0)r+\frac{k}{2}(z^*r^2-z)=0,\label{r1}
\end{align}
where we have assumed the frequency distribution is Lorentzian, i.e. $g(\omega)=\Delta\pi^{-1}/[\Delta^2+(\omega-\omega_0)^2]$. This assumption is necessary to obtain Eq.~(\ref{r1}), however, more generally the ansatz of Ott and Antonsen~\cite{OA1} can be applied to other forms of $g(\omega)$ and treated numerically~\cite{Lafuerza2010PRE}. Eqs.~(\ref{delayed}) and (\ref{r1}) now completely describe the macroscopic oscillator dynamics assuming a fixed coupling strength $k$. We note that Eq.~(\ref{r1}) was derived in Ref.~\cite{Lee1} for time-delayed oscillators without coupling adaptation.

\begin{figure}[t]
\centering
\includegraphics[width=\linewidth]{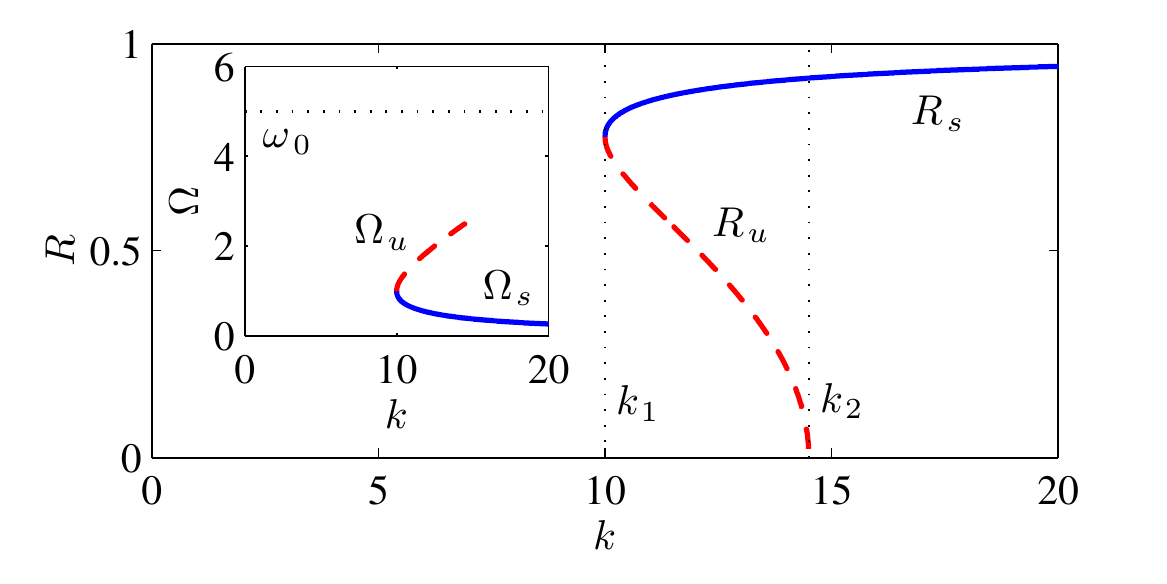}
\caption{(Color online) Solutions $R_s$ (solid blue curve) and $R_u$ (dashed red curve) [given by Eq.~(\ref{eqR})] for the time-delayed system with $\omega_0=5$, $\Delta=1$, and $\tau=1$. Inset: $\Omega_s$ (solid blue curve) and $\Omega_u$ (dashed red curve) corresponding to the angular velocities of the synchronized states. Note that $\Omega_s$ and $\Omega_u$ are much smaller than the average intrinsic frequency $\omega_0=5$ (dotted line).} \label{an}
\end{figure}

Assuming a fixed $k$ value, we now look for steady state solutions by defining $r=Re^{i\psi}$, $z=\rho e^{i\phi}$, and setting $\dot{R}=\dot{\rho}=0$ and $\dot{\psi}=\dot{\phi}=\Omega$. 
Without loss of generality, by rescaling time $t$, the mean natural frequency $\omega_0$, and coupling strength $k$, we can set $\Delta=1$. We also set the time-delay parameter $\tau=1$.
As shown in Fig.~\ref{an} for $\omega_0=5$, in addition to the incoherent solution $R=\rho=0$, a pair of synchronized solutions appear at $k_1=2\omega_0$, given by~\cite{Laing1}
\begin{align}
R_{s/u}&=\frac{\sqrt{\omega_0^2-k \pm \sqrt{k^2-4 \omega_0^2}}}{\omega_0}, \label{eqR}\\
\rho_{s/u}&=\frac{R_{s/u}}{\sqrt{1+\Omega_{s/u}^2}}, \label{eqrho}
\end{align}
with a corresponding angular velocity $\Omega$ given by
\begin{align}\label{eqOmega}
\Omega_{s/u} = \frac{k\mp\sqrt{k^2-4\omega_0^2}}{2\omega_0}.
\end{align}
Subscripts $s/u$ denote whether the solution is stable or unstable, respectively. At $k_2=(\omega_0^2+4)/2$ the unstable synchronized branch merges with the incoherent solution, which becomes unstable for $k>k_2$. Note that for $k$ between $k_1$ and $k_2$ we find bistability since there are both coherent and incoherent solutions that are stable to perturbation (the linear stability of these solutions has been discussed in Refs.~\cite{Lee1,Laing1}). Furthermore, along the synchronized branches $\phi$ lags behind $\psi$ by an angle 
\begin{align}
\psi-\phi=\arcsin\left(\Omega_{s/u}/\sqrt{1+\Omega_{s/u}^2}\right).
\end{align}

We now remark on two aspects of this system not previously discussed in Ref.~\cite{Lee1,Laing1}. First, we note that the average angular velocity $\Omega$ of oscillators in a synchronized state is considerably less than the average intrinsic frequency $\omega_0$ (see inset in Fig.~\ref{an}). This reflects the fact that each individual oscillator is coupled not to the instantaneous mean-field, but to the time-delayed version, slowing down the entire synchronized population. 

Second, whereas the distribution of locked oscillators in the standard Kuramoto model is symmetric about the mean oscillator frequency $\omega_0$, this symmetry is broken by the time delays and as a result the distribution of locked oscillators for oscillators with time-delayed coupling is biased toward oscillators with angular frequencies near $\Omega$. Because $\Omega$ is much smaller than $\omega_0$, this distribution of locked frequencies is typically spread asymmetrically around the mean frequency $\omega_0$.
%
We compute the critical frequencies $\omega_{c,\pm}$ separating phase-locked and drifting oscillators by entering a rotating frame in which synchronized oscillators appear stationary by defining $\Theta_n=\theta_n-\phi$. Here $\Theta_n$ evolves according to $\dot{\Theta}_n=\omega_n-\Omega-k\rho\sin(\Theta_n)$, so that $\Theta_n$ reaches an equilibrium and becomes phase-locked if $|\omega_n-\Omega|\le k\rho$, and otherwise drifts indefinitely. Thus, the critical frequencies that separate the drifting and locked populations are $\omega_{c,\pm}=\Omega\pm k\rho$.

\subsection{Slow coupling adaptation}

Having solved the oscillator dynamics that evolve on the fast time scale, we now study adaptation given by Eq.~(\ref{kdot}) that evolves on a slow time scale. For simplicity, we assume that $k$ relaxes to a linear function of $\rho$,
\begin{equation}\label{eqkev}
G(k,\rho) = \alpha + \beta \rho - k,
\end{equation}
and will study the resulting behavior as a function of the parameters $\alpha$ and $\beta$. While this form for $G$ is not essential, it simplifies our exploration of complex macroscopic behavior under adaptive uniform coupling while yielding rich dynamics. Using Eqs.~(\ref{kdot}) and (\ref{eqrho}), the behavior of the order parameter magnitude $\rho$ and coupling strength $k$ is described on the slow time scale by the ODE
\begin{align}\label{topbranch}
T\dot{k} = \alpha + \beta  \rho_s(k) - k,
\end{align}
when the system is synchronized, and
\begin{align}\label{lowbranch}
T\dot{k} = \alpha  - k,
\end{align}
when the system is incoherent (i.e., $\rho=0$). 

As shown in Fig.~\ref{inter}(a), when the system is in the incoherent state and $k$ surpasses $k_2$ a dynamic bifurcation occurs in a rapid transition from incoherence to synchronization. Similarly, when the system is in the synchronized state and $k$ decreases below $k_1$ another dynamic bifurcation occurs in a rapid transition from synchronization to incoherence. These rapid transitions from one branch to the other represent discontinuous phase transitions, which have also been refered to as explosive synchronization~\cite{Explosive}. Again, $T$ is assumed to be large enough that state transitions occur with fixed $k$. Furthermore, we assume that, upon a perturbation of the oscillator phases which changes the value of $\rho$ and $R$, the system returns to the inertial manifold in which the Ott-Antonsen ansatz is valid on a time scale which is much faster than $T$, so that we can assume $k$ is constant during this process. It follows that the macroscopic behavior depends on the location of the stable fixed points of Eqs.~(\ref{topbranch}) and (\ref{lowbranch}) as depicted for various situations in Figs.~\ref{inter}(a)-\ref{inter}(d). 

We now classify the nature of macroscopic behavior by studying the stable fixed points of Eq.~(\ref{topbranch}) (the {\it synchronized fixed point}, $k^*_{sync}$) and Eq.~(\ref{lowbranch}) (the {\it incoherent fixed point}, $k^*_{inc}$). Note that incoherent fixed points for $k>k_2$ are not relevant since the incoherent branch is unstable in that region, and therefore we will ignore these fixed points in what follows. Classifying the macroscopic dynamics for a particular choice of $(\alpha,\beta)$ reduces to the analysis of the stable fixed points of Eqs.~(\ref{topbranch}) and (\ref{lowbranch}) subject to the incoherent solution ($\rho=0$) and stable synchronized solution [Eqs.~(\ref{eqrho})]. The different possible macroscopic behaviors are the following:

\begin{figure*}[t]
\centering
\includegraphics[width=\linewidth]{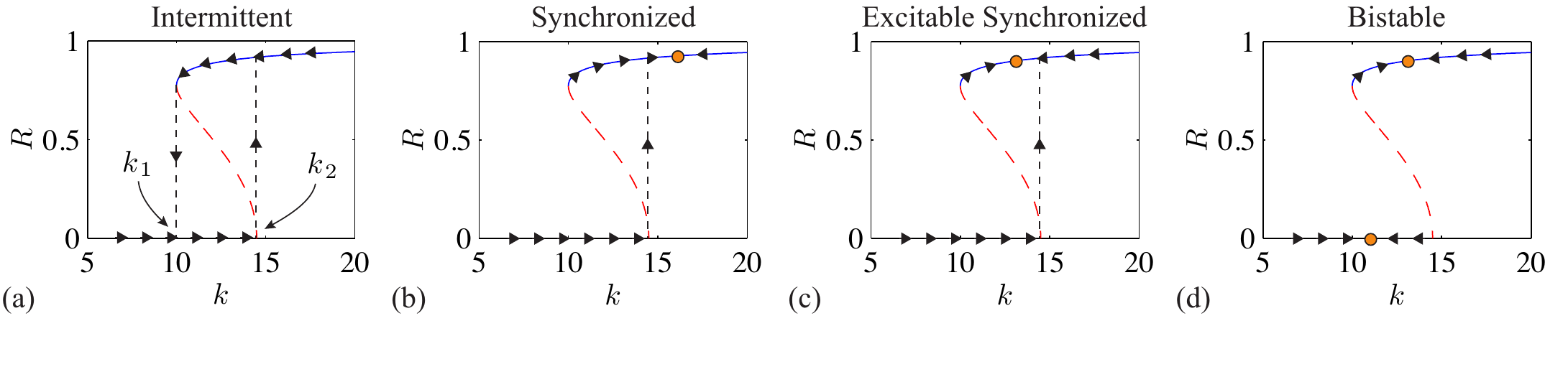}
\caption{(Color online) Various macroscopic behaviors may occur for uniform adaptation following Eq.~(\ref{eqkev}) depending on $\alpha,\beta$ and the location of fixed points (circles). Examples include (a) intermittent, (b) synchronized, (c) ES, and (d) bistable states.} 
\label{inter}
\end{figure*}

\begin{itemize}

\item {\bf No synchronized fixed point, no incoherent fixed point.} As shown in Fig.~\ref{inter}(a), if no stable fixed point exists on either branch the system will repeat the following macroscopic oscillation: $k$ will increase along the incoherent branch, then after the system synchronizes at $k_2$, it will decrease along the synchronized branch, until desynchronization occurs at $k_1$. We define this behavior as the {\it intermittent} state and investigate its properties in Sec.~IIC.
\item {\bf Synchronized fixed point, no incoherent fixed point.} If a stable fixed point occurs only on the synchronized branch, we define two subclasses for this state: if $k^*_{sync}\ge k_2$ we define the macroscopic state as the {\it synchronized} state [see Fig. \ref{inter}(b)], whereas if $k^*_{sync}<k_2$ we refer to the state as the {\it excitable synchronized} (ES) state [see Fig. \ref{inter}(c)]. This distinction is made to account for the possibility that perturbations to the value of $\rho$ in the ES state (e.g., due to noise or finite-size fluctuations~\cite{FS}) may desynchronize the system by decreasing $\rho$ below $\rho_u$, which would result in $\rho\to0$ followed by $k$ increasing until $k=k_2$, after which the system will synchronize and return to the fixed point. Thus, the synchronized state can be interpreted as the resting state of an excitable system and a temporary desynchronization as an excitation.
\item {\bf Incoherent fixed point, no synchronized fixed point.} Analogous to the previous case, we define two sub-classes for this state: if $k^*_{inc}<k_1$ we refer to this state as the {\it incoherent} state, whereas if $k^*_{inc}\ge k_1$ we refer to this state as the {\it excitable incoherent} (EI) state. Again this distinction is made to account for the possibility of perturbations to the value of $\rho$ in the EI state that can produce a temporary synchronization. In this case the system can be synchronized if $\rho$ is increased above $\rho_u$, resulting in $\rho\to\rho_s$ followed by $k$ decreasing until $k=k_1$, desynchronization, and finally a return to the fixed point. In this scenario, the incoherent fixed point can be interpreted as the resting state of an excitable system and a temporary synchronization as an excitation.
\item {\bf Synchronized fixed point, incoherent fixed point.} If stable fixed points occur on both branches, we refer to this state as the {\it bistable} state, an example of which is shown in Fig.~\ref{inter}(d).
\end{itemize}

\begin{figure}[t]
\centering
\includegraphics[width=0.8\linewidth]{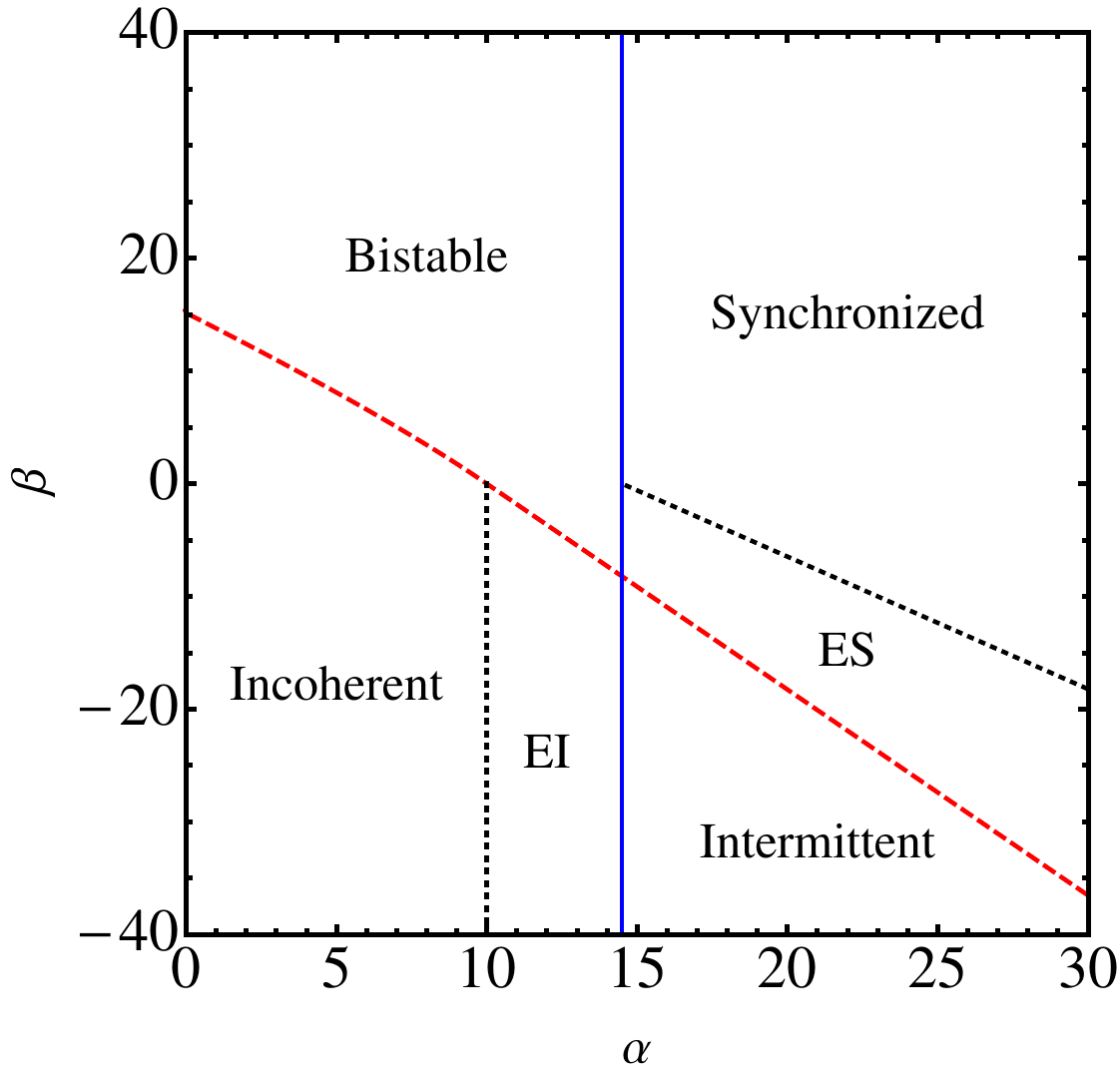}
\caption{(Color online) Bifurcation diagram summarizing boundaries between intermittent, synchronized, ES, incoherent, EI, and bistable states for $\omega_0=5$, $\Delta=1$, and $\tau=1$.} 
\label{phasespace}
\end{figure}

To find the location of the bifurcations between these states, we calculate the critical $\alpha,\beta$ that correspond to the formation or destruction of fixed points on either branch. For fixed points on the incoherent branch, this occurs at $\alpha=k_2$. For fixed points on the synchronized branch, we require that the curves $k-\alpha$ and $\beta\rho_s(k)$ are tangent if $\beta\ge0$, which occurs when
\begin{equation}
\beta \rho_s(k) = k - \alpha,~~ \beta  \frac{d\rho_s(k)}{dk} = 1,
\end{equation}
 and coincide if $\beta<0$, which happens when
\begin{equation}
\alpha + \beta \rho_s(k_1) = k_1.
\end{equation}
Finally, the boundary between EI and incoherent states is given by $\alpha=k_1$ (the incoherent fixed point entering the bistable region), while the boundary between ES and synchronized states is given by the curve $\alpha+\beta\rho_s(k_2)=k_2$ (the synchronized fixed point entering the bistable region). In Fig.~\ref{phasespace} we show the bifurcation diagram for $\omega_0=5$ and $\Delta=1$ by plotting curves describing the formation/destruction of incoherent fixed points in solid blue, synchronized fixed points in dashed red, and the borders between EI/ES and Incoherent/Synchronized states in dotted black. We label regions with the states described above. We note that excitable and intermittent states are possible only when $\beta < 0$, which we refer to as {\it anti-Hebbian} adaptation (accordingly, we refer to $\beta>0$ as Hebbian adaptation). This terminology is based on the observation that for $\beta>0$ ($\beta<0$) in Eq.~(\ref{eqkev}), coupling is promoted (inhibited) by the synchrony of oscillators.

\subsection{Intermittent case}

\begin{figure*}[t]
\centering 
\includegraphics[width=0.48\linewidth]{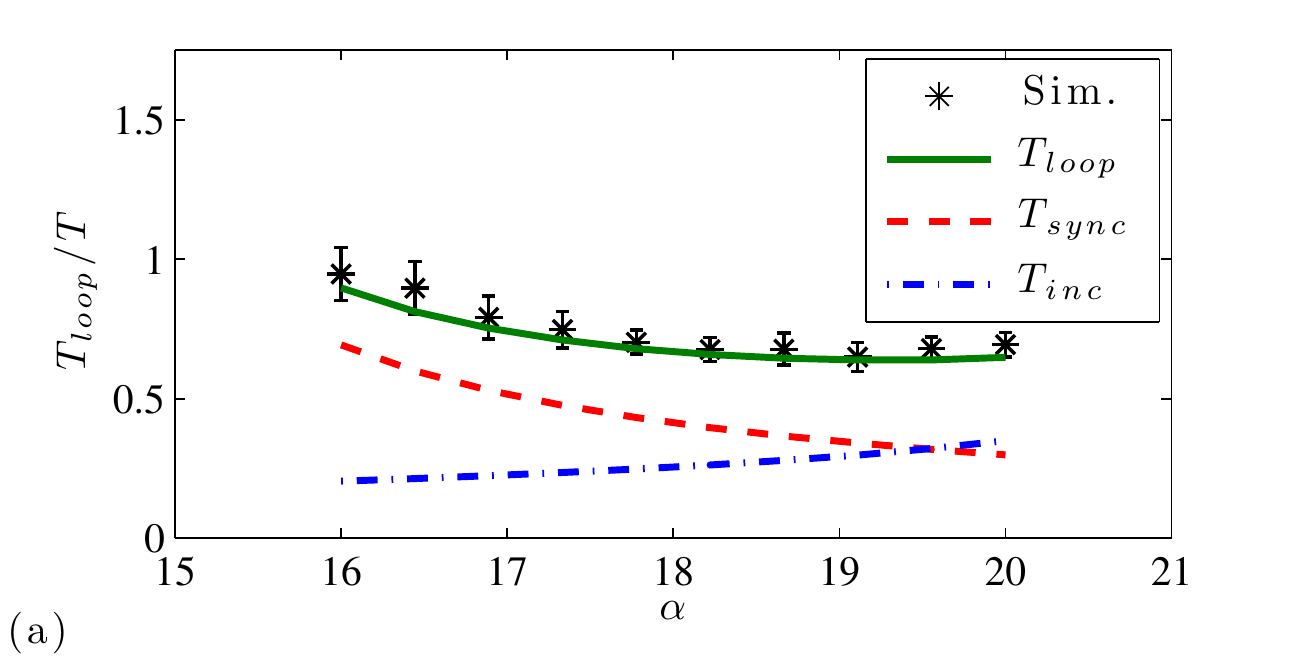}
\includegraphics[width=0.48\linewidth]{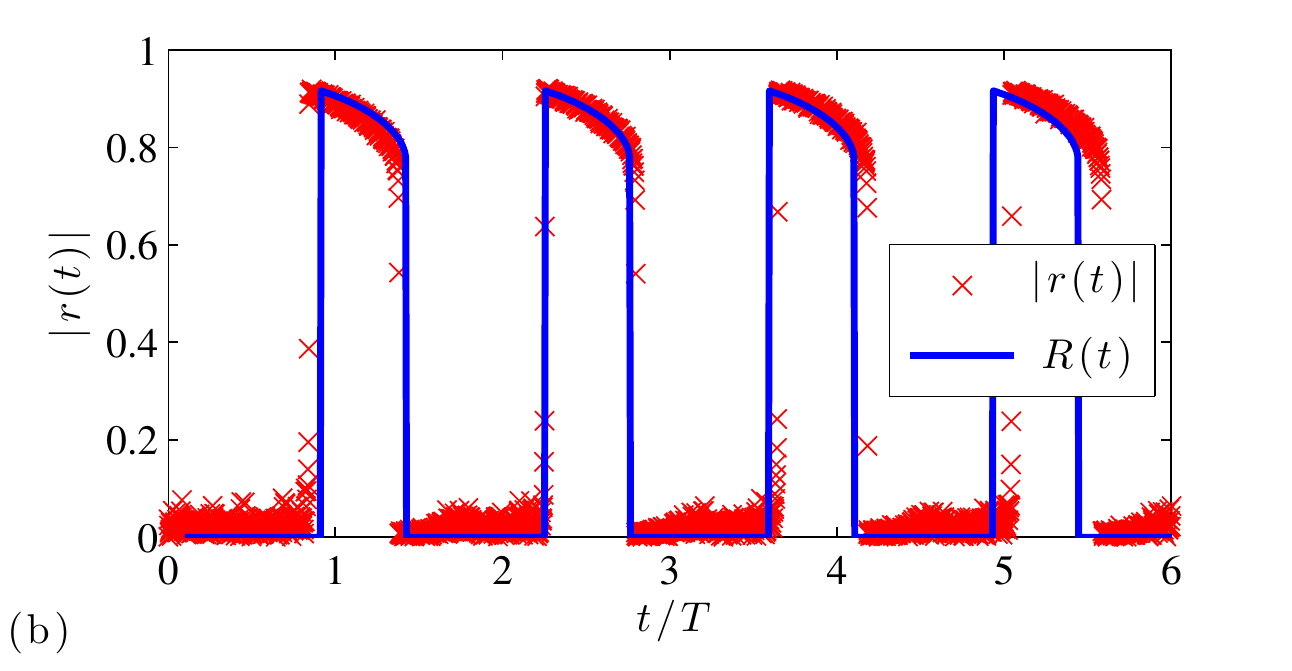}
\caption{(Color online) (a) Theoretical $T_{loop}$ (solid green), $T_{inc}$ (dashed red), and $T_{sync}$ (dot-dashed blue) agree well with $T_{loop}$ averaged over 16 simulations (black asterisks, where error bars indicate standard deviation). Other parameters are $\beta=-20$, $\omega_0=5$, $\Delta=1$, $\tau=1$, $T=2000$, and $N=5000$. (b) For $\alpha = 18$ theoretical values for $R(t)$ (solid blue) agree well with direct simulation $|r(t)|$ (red crosses).}
\label{T}
\end{figure*}

Motivated by observations of intermittently synchronous dynamics in various applications of oscillator systems (e.g., in neural activity \cite{Deco1,Park1} and clapping audiences \cite{clapping}) we now study in detail the intermittent case illustrated in Fig.~\ref{inter}(a) and characterized by intermittent periods of macroscopic synchronization. Of interest is the period of oscillation, which can be found by integrating the time spent following the incoherent and synchronized branches of the bistable region. The time spent in the incoherent state, $T_{inc}$, corresponds to the time it takes for $k$ to increase from $k_1$ to $k_2$ with $\rho = 0$ and is given by
\begin{align}\label{tsync}
T_{inc} = T \ln\left(\frac{\alpha -k_1}{\alpha -k_2}\right).
\end{align}
Similarly, the time spent in the synchronized state, $T_{sync}$, corresponds to the time it takes for $k$ to decrease from $k_2$ to $k_1$ along the synchronized branch and is given by
\begin{align}\label{tdesync}
T_{sync} = T\int_{k_2}^{k_1}\frac{dk}{\alpha + \beta \rho_s(k) -k}.
\end{align}
Since we assume that the timescale of adaptation is much larger than the timescale of oscillator dynamics, we neglect the time it takes for oscillators to synchronize and desynchronize at $k_2$ and $k_1$, respectively. This gives the period of oscillation $T_{loop} = T_{sync}+T_{inc}$. Fig.~\ref{T}(a) shows $T_{loop}$ (solid green curve), $T_{inc}$ (dashed red curve), and $T_{sync}$ (dot-dashed blue curve) as a function of $\alpha$ for $\omega_0=5$, $\Delta=1$, and $\beta = -20$. For these parameters $k_1=10$ and $k_2=14.5$. In addition, we compute the period of oscillation from simulating $N=5000$ oscillators with $T=2000$, plotting the mean of $T_{loop}$ over 16 simulations at each $\alpha$ (black asterisks). Error bars indicate the standard deviation. While $T_{inc}$ and $T_{loop}$ diverge as $\alpha \to k_2^{-}$, $T_{sync}$ and $T_{loop}$ remain finite as $\alpha \to [k_1 - \beta \rho_s(k_1)]^{+}$, since the square root singularity of $\rho_s(k)$ at $k = k_1$ prevents the integral in Eq.~(\ref{tdesync}) from diverging.

As shown in Fig. \ref{T}(b), the macroscopic behavior of the system oscillating between incoherent and synchronized states may be described by considering the low dimensional system given by Eqs.~(\ref{eqR}), (\ref{eqrho}), (\ref{topbranch}), and (\ref{lowbranch}). This theoretical solution $R(t)$ (solid blue curve) agrees well with the order parameter's magnitude $|r(t)|$ (red crosses) from direct simulation of the high-dimensional system given by Eqs.~(\ref{thetadot}), (\ref{delayed}), (\ref{kdot}), and (\ref{eqkev}). The simulation in Fig.~\ref{T}(b) was done with $N = 5000$ oscillators with $T = 2000$, $\alpha = 18$, and $\beta = -20$. Remarkably, the behavior of the high-dimensional system is captured well by this piecewise defined one-dimensional ODE. The period taken from simulations is slightly longer than our theoretical solution, which is most likely due to two effects. First, our theoretical solution neglects the synchronization and desynchronization times at the dynamic bifurcations occurring at $k=k_1$ and $k_2$. Second, along the incoherent branch the value of the order parameter in simulations typically takes values of size $~\mathcal{O}(N^{-1/2})$~\cite{FS} rather than zero, which slightly slows down the adaptation.

\subsection{Excitable incoherent case}

\begin{figure}[t]
\centering 
\includegraphics[width=\linewidth]{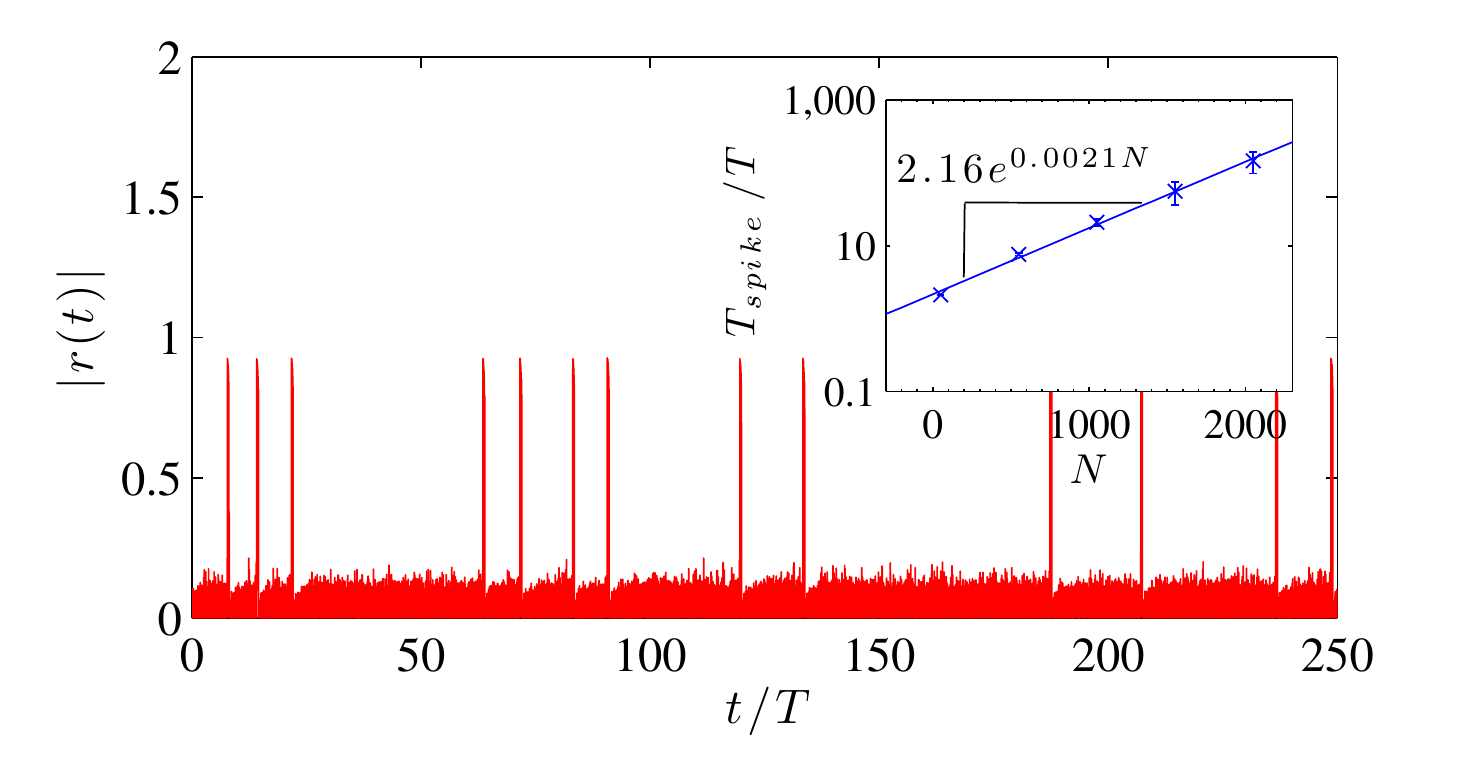}
\caption{(Color online) Spiking events are shown for the order parameter $r$ when $\alpha=14.45$ and $\beta=-20$ are chosen so that our system is in the EI state. Note that the timescale of spiking is dominated by the spontaneous synchronization process \cite{dane}. (inset) For a system of size $N$ we may predict the expected time between spikes as $T_{spike} \varpropto \exp(\zeta N)$ for some constant $\zeta$ \cite{dane}. The solid line indicates a least-squares fit $T_{spike}/T=2.16\exp(0.0021N)$.}
\label{fig:spike}
\end{figure}

We conclude our analysis of this model by studying the EI state. As previously mentioned, if the system is in the incoherent state with $k=k^*_{inc}$, a perturbation to the order parameter can  cause $r$ to become larger than the unstable solution $r_u$, resulting in a dynamic bifurcation. While $k$ remains fixed during this rapid transition, after synchrony $k$ will evolve until the system returns to the equilibrium of $r\sim0$ and $k=k^*_{inc}$. In particular, for finite systems this perturbation could occur due to finite size effects, resulting in a spontaneous synchronization event~\cite{dane}. This can be viewed as a random spiking event for the macroscopic dynamics, which corresponds to the oscillators synchronizing very briefly relative to the typical time between spikes. 

Spiking events are shown in Fig. \ref{fig:spike}, where we plot $|r(t)|$ versus time for $\tau=1$, $T=1000$, $\Delta=1$, $\omega_0=5$, and $N=1050$. Note that the system spends the majority of time in the incoherent state, and the slow timescale of spontaneous synchronization dominates other time scales. Defining the average time between synchronization events as the {\it inter-spike time}, $T_{spike}$, we briefly discuss the dependence of $T_{spike}$ on system size $N$. In Ref. \cite{dane} it was shown that the spontaneous synchronization event can be modeled as a Kramer escape process where the expected escape time is proportional to $\exp (\zeta N)$ for some constant $\zeta $. Therefore, because the escape process dominates the timescale of dynamics, we expect that the inter-spike time scales as $T_{spike} \propto \exp (\zeta N)$. This is confirmed in the inset of Fig. \ref{fig:spike}, where $T_{spike}$ is shown to vary exponentially with $N$. The solid line is a least-squares fit $T_{spike}/T=2.16\exp(0.0021N)$.

\section{Other models}

In the previous section we analyzed in detail the model given by Eqs.~(\ref{thetadot})-(\ref{kdot}), which describes a system of oscillators with heterogeneous natural frequencies and heterogeneous time-delays subject to uniform coupling adaptation~\cite{Lee1}. The purpose of this model was to illustrate generic behavior occurring in adaptive networks with bistable regimes. In this section we study numerically and analytically several other models which have been selected to show that the type of behaviors observed in the previous section occur more generally. In particular, in Sec.~3.1 we investigate network adaptation, which is often used in Kuramoto-type models of information processing and memory in neural networks~\cite{STDP,Hebbian}. In Sec.~3.2 we explore complex macroscopic behavior that can arise for adaptation in networks containing community structure. Finally, in Sec. 3.3 we show that our findings apply to other oscillator systems exhibiting multistability (e.g., due to frequency adaptation \cite{dane} or inertia \cite{inertia}) by studying adaptation in oscillator systems with bistability due to a bimodal distribution of intrinsic frequencies \cite{Martens2}.

\subsection{Network adaptation}

\begin{figure}[t]
\centering
\includegraphics[width=\linewidth]{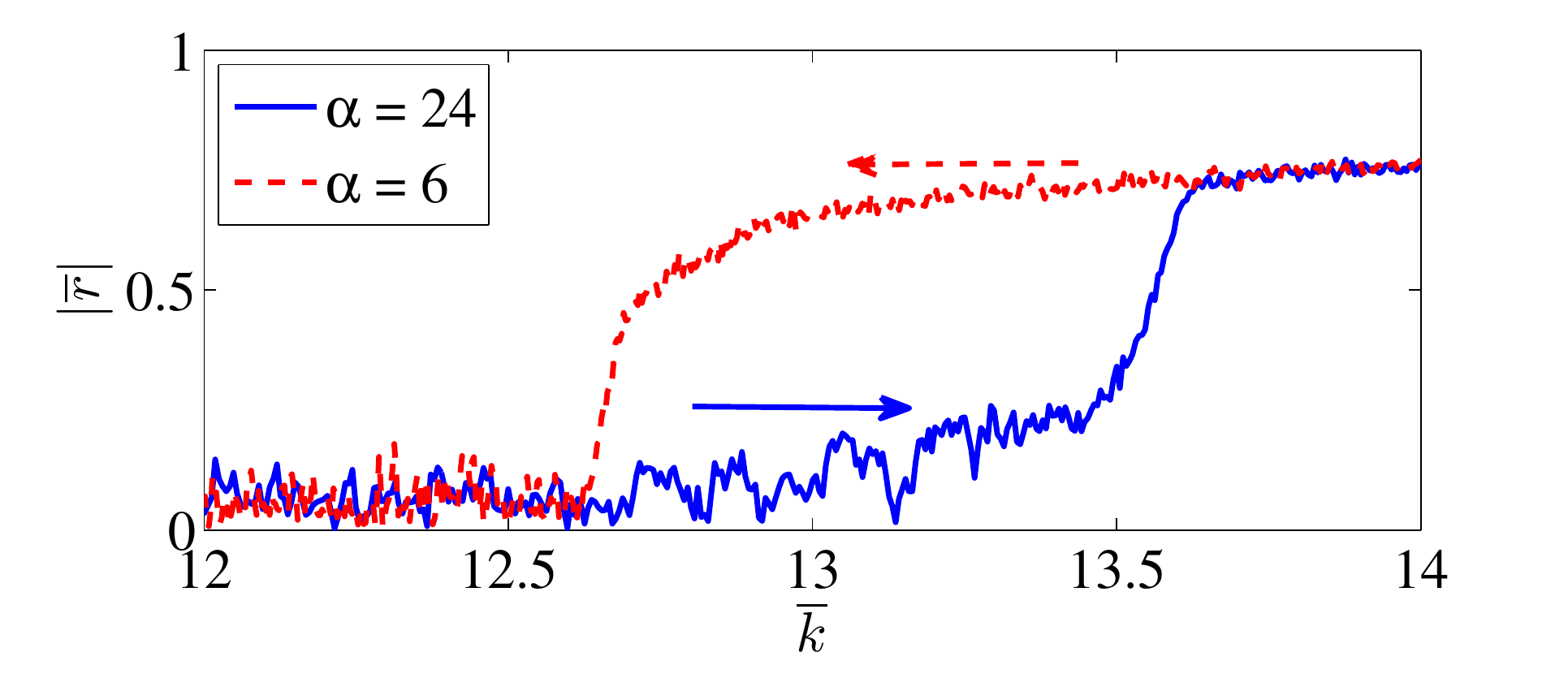}
\caption{(Color online) Example $(\overline{k},|\overline{r}|)$ trajectories of the system given by Eqs.~(\ref{thetadotA})-(\ref{kdotA}). The solid blue and dashed red trajectories were obtained using $(\alpha,\beta)=(24,0)$ and $(6,0)$, respectively with an initial coupling strengths $\overline{k}=6$ and $22$, respectively. Other parameters are $\omega_0=5$, $\Delta=1$, $\tau=1$, $T=1000$, $\gamma=3$, $d_{min}=100$, and $N=1000$.} \label{netExample}
\end{figure} 

First, we will consider a system similar to Eqs.~(\ref{thetadot})-(\ref{kdot}) in which the interactions between oscillators are not mediated by a global mean field, but occur instead through an underlying coupling network. We assume the undirected network structure is represented by an adjacency matrix $A$, where
\begin{align}
A_{nm} = \left\{\begin{array}{rl}1&\text{ if a link exists from oscillator $m$ to oscillator $n$,}\\0 &\text{ if no link exists.}\end{array}\right.\nonumber
\end{align}
Introducing a coupling weight $k_{nm}$ to each link and using the locally-defined order parameters $r_n$, where
\begin{align}
r_n=\sum_{m=1}^NA_{nm}k_{nm}e^{i\theta_m},
\end{align}
we consider the system given by
\begin{align}
\dot{\theta}_n&=\omega_n+\lambda_D^{-1}\mbox{Im}(z_ne^{-i\theta_n}), \label{thetadotA}\\
\tau \dot{z}_n&=r_n-z_n, \\
T\dot{k}_{nm}&=\alpha+\beta Re(r_nz_n^*)-k_{nm},\label{kdotA}
\end{align}
where $\omega_n$ is again randomly drawn from a Lorentzian with mean $\omega_0$ and spread $\Delta$, and $\lambda_D$ is the dominant eigenvalue of $A$. We normalize the coupling term in Eq.~(\ref{thetadotA}) by $\lambda_D$ so that the $k_{nm}$ values producing bistability are on the same order as $k$ values that yield bistability in the uniform adaptation model~\cite{JGR1}. To measure the global degree of synchrony and coupling strength we introduce the average order parameter
\begin{align}
\overline{r}=\frac{\sum_{n} r_n}{\sum_{n,m}A_{nm}k_{nm}}\in[0,1],
\end{align}
and average coupling strength
\begin{equation}
\overline{k}=\frac{\sum_{n,m}A_{nm}k_{nm}}{\sum_{n,m}A_{nm}}.
\end{equation}

Since we are assuming that oscillator $n$ is affected by a delayed order parameter, the adaptation of the coupling $k_{nm}$ between oscillators $m$ and $n$, Eq.~(\ref{kdotA}), is assumed to depend on the local instantaneous order parameter of oscillator $n$, $r_n$, and the  delayed order parameter at oscillator $m$, $z_m$. As before, we interpret positive values of $\beta$ as Hebbian adaptation, and negative values as anti-Hebbian adaptation.

\begin{figure}[t]
\centering
\includegraphics[width=0.8\linewidth]{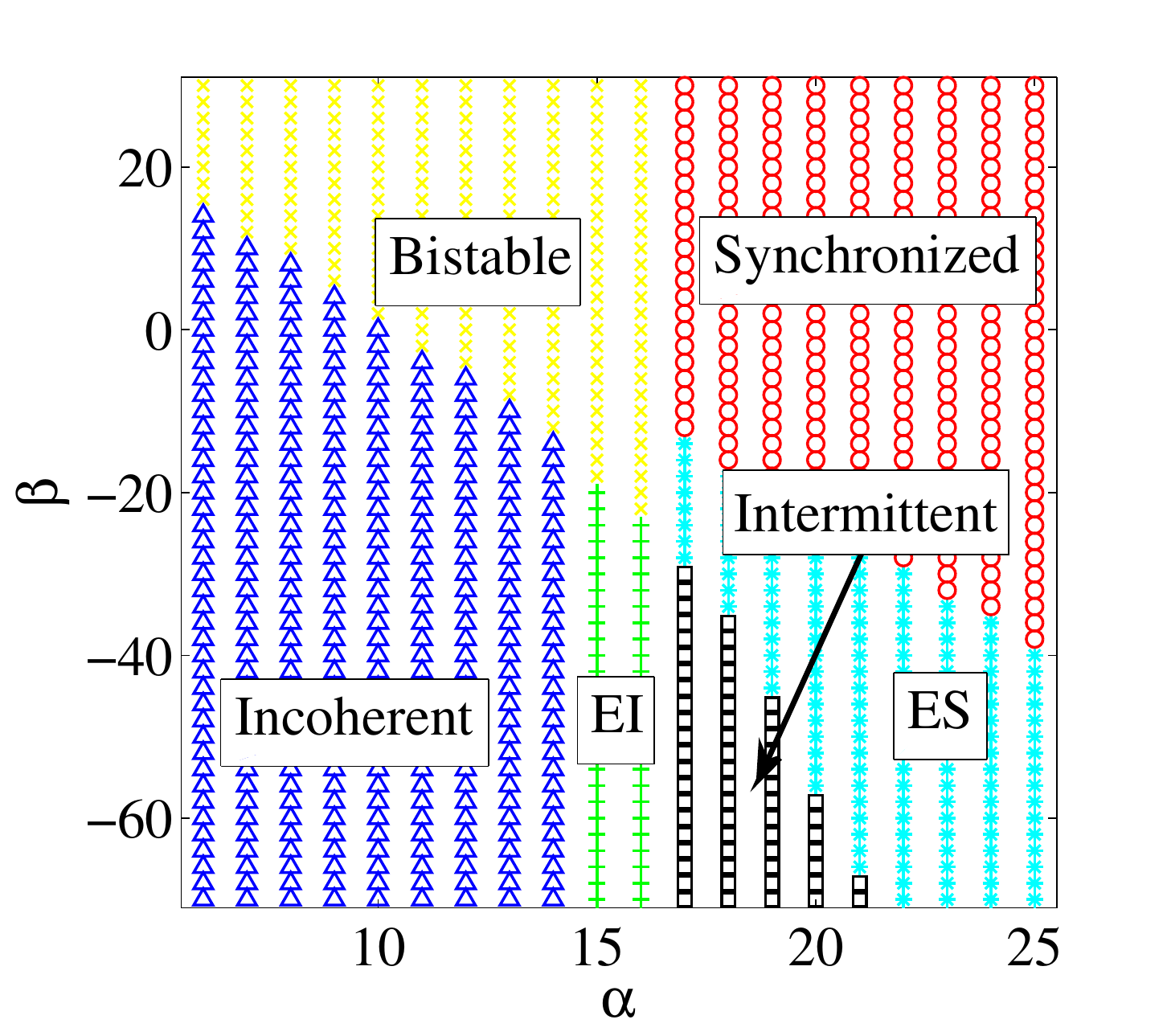}
\caption{(Color online) Bifurcation diagram summarizing oscillatory (black squares), synchronized (red circles), ES (cyan asterisks), incoherent (blue triangles), EI (green plusses), and bistable (yellow crosses) states for network adaptation of time-delayed oscillators with $\Delta=1$, $\omega_0=5$, $\tau=1$, and $T = 1000$.}
\label{netplane}
\end{figure}

\begin{figure*}[t]
\centering
\includegraphics[width=0.45\linewidth]{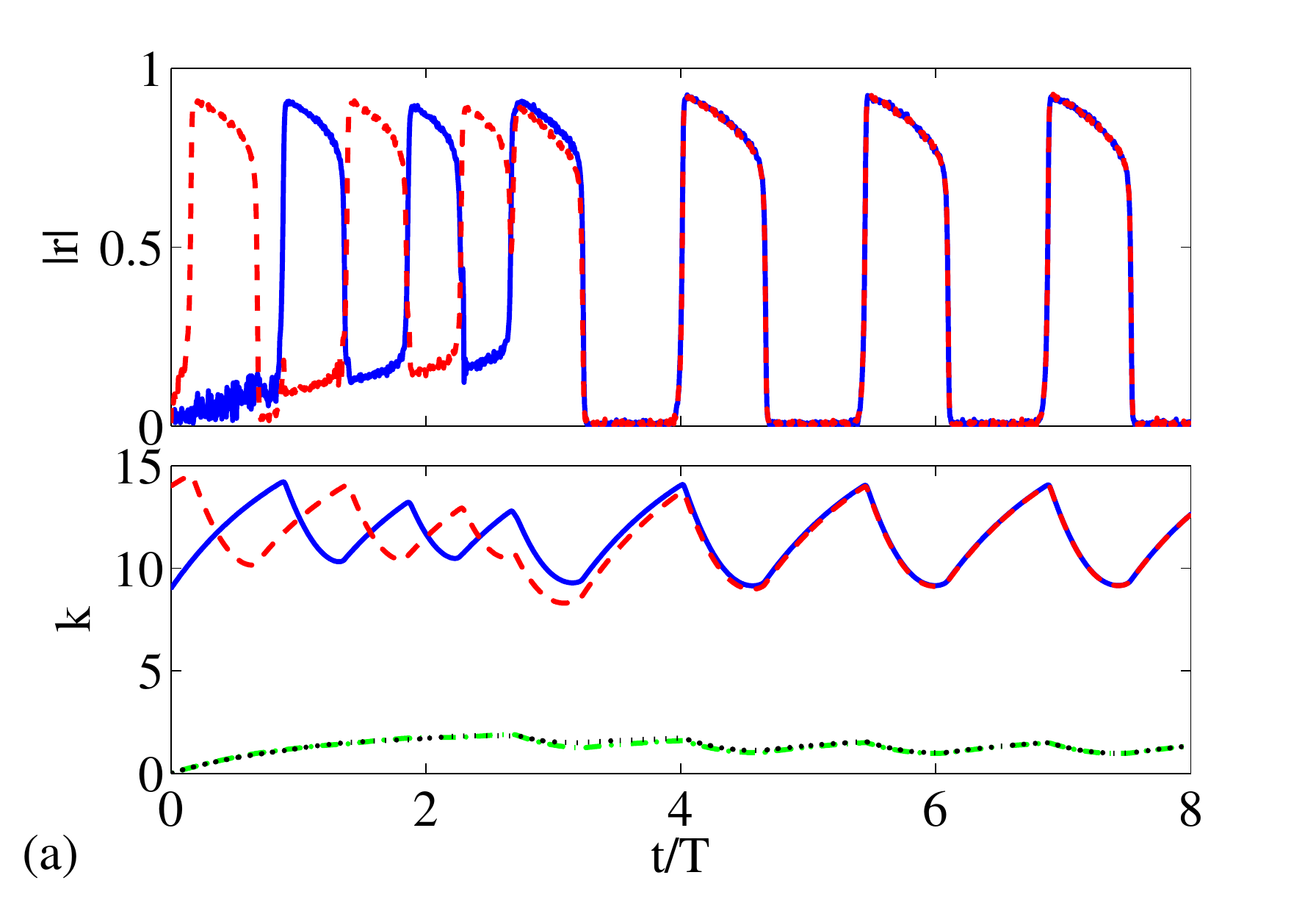}
\includegraphics[width=0.45\linewidth]{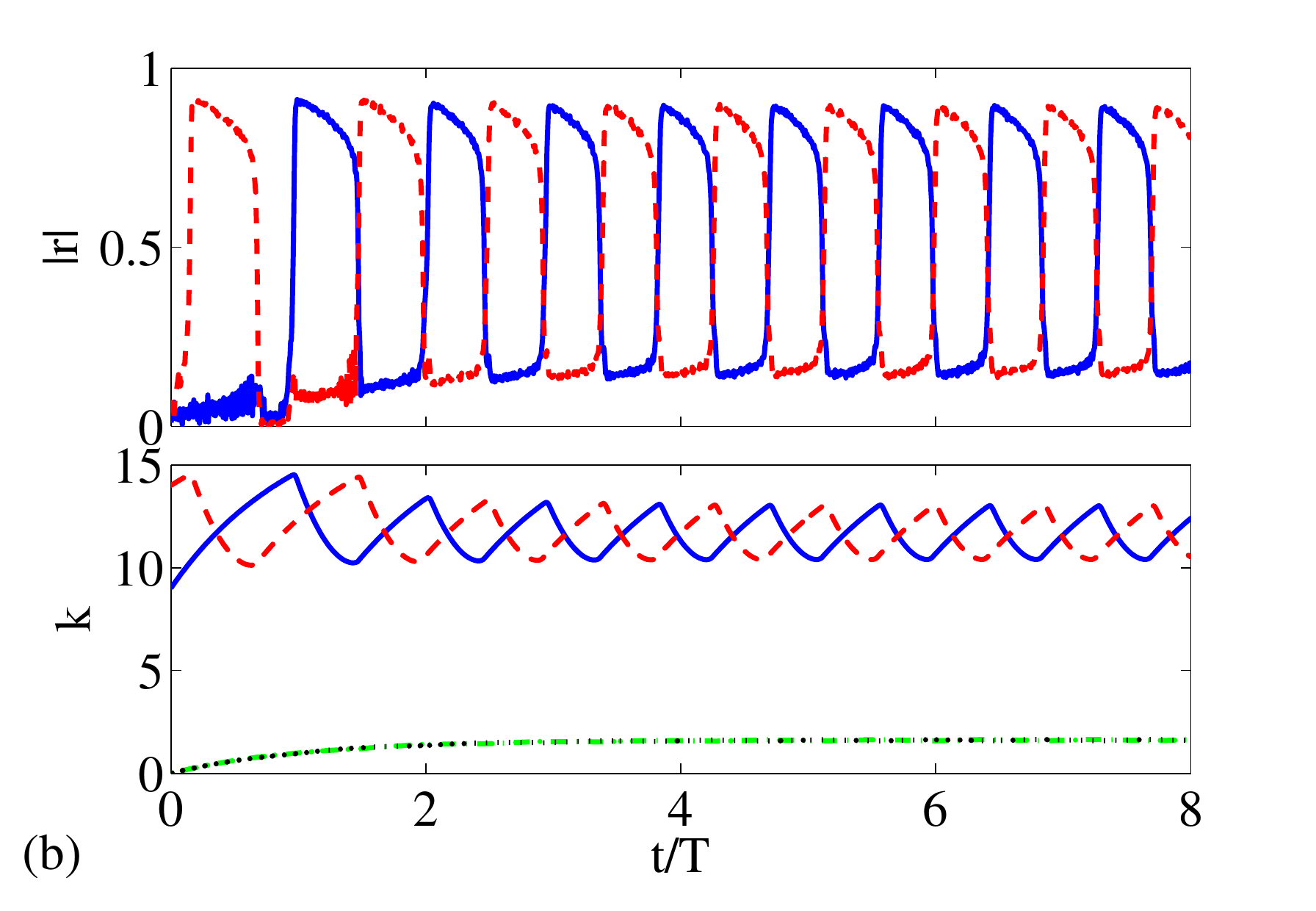}
\caption{(Color online) Community interaction model with parameters $\omega_0=5$, $\Delta=1$, $\tau=1$, $\alpha=18$, $\beta=-30$, $T=200$, and $\epsilon=0.105$ (a) and $0.085$ (b). Top panels: evolution of $|r_{1}|$ (solid blue line) and $|r_{2}|$ (dashed red line), bottom panels: evolution of $k_{11}$ (solid blue line), $k_{22}$ (dashed red line), $k_{12}$ (dot-dashed green line), and $k_{21}$ (dotted black line).} \label{comms}
\end{figure*} 

Using the Chung-Lu model~\cite{ChungLu}, we construct an undirected network with a power-law degree distribution, $P(d)\propto d^{-\gamma}$, with exponent $\gamma=3$, minimum degree $d_{min}=100$, and $N=1000$ oscillators, where the degree $d$ is defined as $d_n=\sum_{m=1}^NA_{nm}$. The parameters for the oscillator dynamics are $\omega_0=5$, $\Delta=1$, $\tau=1$, and the adaptive timescale is $T=1000$. The dominant eigenvalue for the network constructed for the simulations shown here is $\lambda_D=232.325$. In Fig.~\ref{netExample} we show representative $(\overline{k},|\overline{r}|)$ trajectories. First using $(\alpha,\beta)=(24,0)$, we allow the average coupling strength to increase from an initial value of $\overline{k}=6$ (solid blue trajectory). Next using $(\alpha,\beta)=(6,0)$, we allow $\overline{k}$ to decrease from an initial value of $\overline{k}=22$ (dashed red). We find that in analogy with the uniform adaptation case, a stable synchronized solution $|R|>0$ is created at $\overline{k}=\overline{k}_1\approx 12.6$, and the incoherent solution $|R|=0$ becomes unstable at $\overline{k}=\overline{k}_2\approx13.6$. Thus, dynamic bifurcations occur approximately when an incoherent state's average coupling increases through $\overline{k}_2$ or a synchronized state's average coupling decreases through $\overline{k}_1$.

Next we numerically explore the $(\alpha,\beta)$ parameter space, classifying the observed behaviors as bistable, intermittent, synchronized, ES, incoherent, and EI, following the criteria in Sec. 2. In Fig.~\ref{netplane} we plot the results. Starting from the top left and proceeding clock-wise, we plot bistable (yellow crosses), synchronized (red circles), ES (cyan asterisks), oscillatory (black squares), EI (green plusses), and incoherent (blue triangles). These states were found by tracking the trajectories of $|\overline{r}|$ and $\overline{k}$ for two simulations at each pair $(\alpha,\beta)$, one trajectory starting from an incoherent state with $\overline{k}<\overline{k}_1$, and the other starting from a synchronized state with $\overline{k}>\overline{k}_2$. The results are smooth enough so that boundaries between regions are clear. As expected, while the exact boundaries in Fig.~\ref{netplane} differ from those plotted in Fig.~\ref{phasespace}, the topologies of the two phase spaces agree qualitatively.

\subsection{Community interaction}

Next, we generalize the system studied in Sec. 2 to a two community model where coupling is strong within communities and weak between communities. For simplicity, we assume adaptation within and between each community is uniform. The model we consider is:
\begin{align}
\dot{\theta}_n^\sigma &= \omega_n^\sigma + \sum_{\sigma'=1}^2k_{\sigma\sigma'}\text{Im}(z_{\sigma'}e^{-i\theta_n^\sigma}), \\
\tau_\sigma\dot{z}_\sigma &=r_\sigma-z_\sigma, \label{eqcomz}\\
T\dot{k}_{\sigma\sigma'} &=G^{\sigma\sigma'}(\vec{k},\vec{r},\vec{z}),
\end{align}
where $\sigma=1,2$ denotes the community, $\theta_n^\sigma$ denotes the phase of an oscillator in community $\sigma$, $r_\sigma=\frac{1}{N_\sigma}\sum_{m=1}^{N_\sigma}e^{i\theta_m^\sigma}$ is the Kuramoto order parameter over oscillators in community $\sigma$, $\vec{k} = [k_{11},k_{12},k_{21},k_{22}]^T$, $\vec{r}=[r_1,r_2]^T$, $\vec{z}=[z_1,z_2]^T$, and the natural frequencies $\omega_n^\sigma$ are drawn from the distribution $g_\sigma(\omega)$. 

Separating the fast oscillator dynamics from the slow adaptation dynamics as before, a dimensionality reduction for the $N_\sigma\to\infty$ limit as in Refs.~\cite{seb2,OA1} yields
\begin{align}
\dot{r}_\sigma&=(-\Delta_\sigma+i\omega_0^\sigma)r_\sigma+\frac{1}{2}\sum_{\sigma'=1}^2k_{\sigma\sigma'}(z_{\sigma'}-z_{\sigma'}^*r_\sigma^2),\label{eqcomr}
\end{align}
where we have assumed that the distribution $g_\sigma(\omega)$ is Lorentzian with spread $\Delta_\sigma$ and mean $\omega_0^\sigma$. Eqs.~(\ref{eqcomr}) and (\ref{eqcomz}) give the low-dimensional evolution of oscillator dynamics. Furthermore, we consider the adaptation dynamics given by
\begin{align}
T\dot{k}_{\sigma\sigma'} = \alpha_{\sigma\sigma'} + \beta_{\sigma\sigma'}\text{Re}(r_\sigma z_{\sigma'}) - k_{\sigma\sigma'}.
\end{align}

Depending on the choices of $\Delta_\sigma$, $\omega_0^\sigma$, $\alpha_{\sigma\sigma'}$, and $\beta_{\sigma\sigma'}$, the resulting dynamics can vary greatly. For simplicity we choose $\Delta_\sigma=\Delta=1$ and $\omega_0^\sigma=\omega_0=5$, $\tau_\sigma=\tau=1$, $\alpha_{\sigma\sigma'}=\alpha$ and $\beta_{\sigma\sigma'}=\beta$ for $\sigma=\sigma'$, and $\alpha_{\sigma\sigma'}=\epsilon\alpha$ and $\beta_{\sigma\sigma'}=\epsilon\beta$ for $\sigma\ne\sigma'$ where $0<\epsilon<1$. We induce oscillatory behavior by choosing $\alpha=18$, $\beta=-30$, and $T=200$, and investigate the effect of varying $\epsilon$. Particularly, we are interested in macroscopic synchrony of the two communities.

\begin{figure*}[t]
\centering
\includegraphics[width=0.48\linewidth]{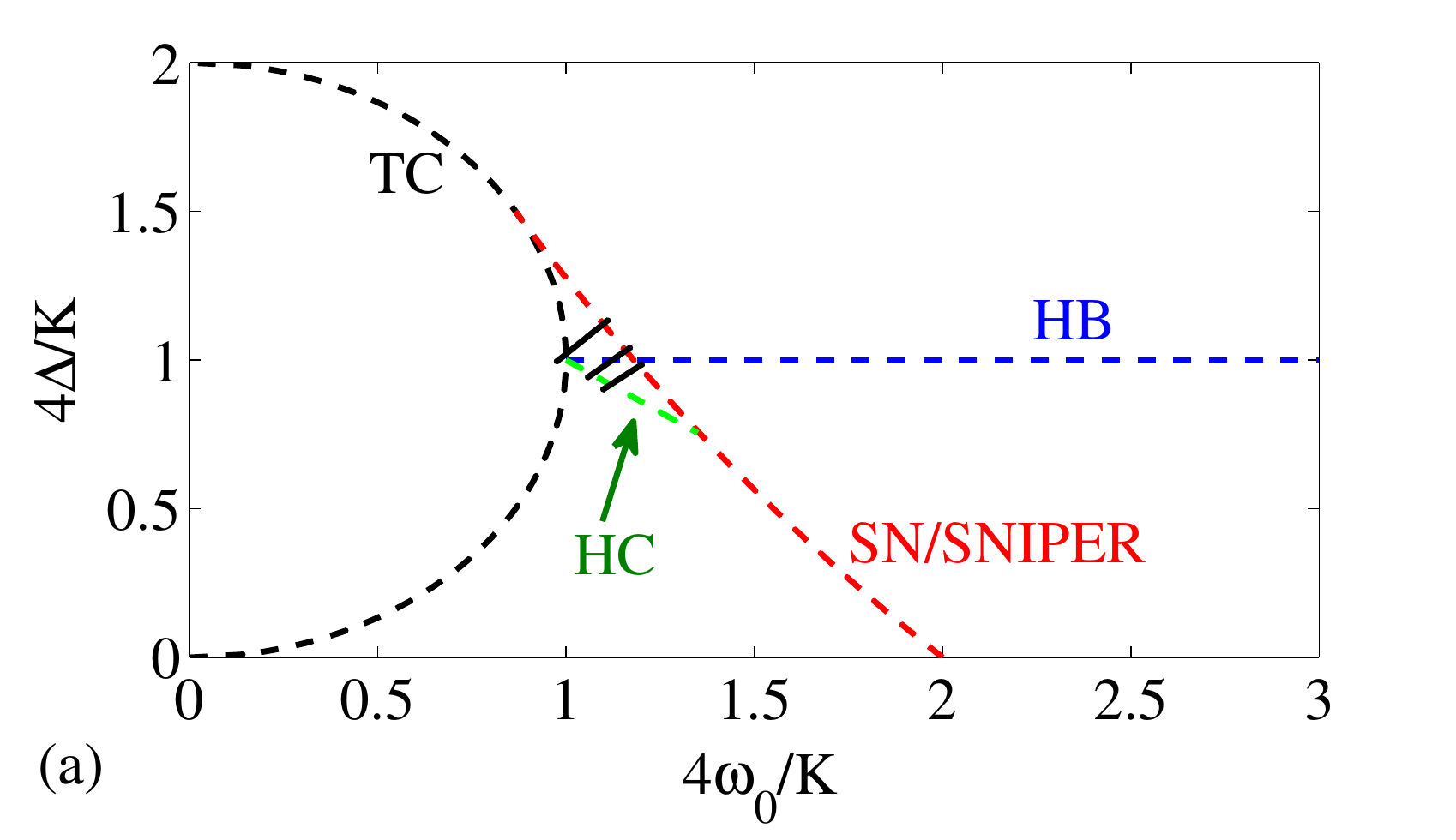}
\includegraphics[width=0.48\linewidth]{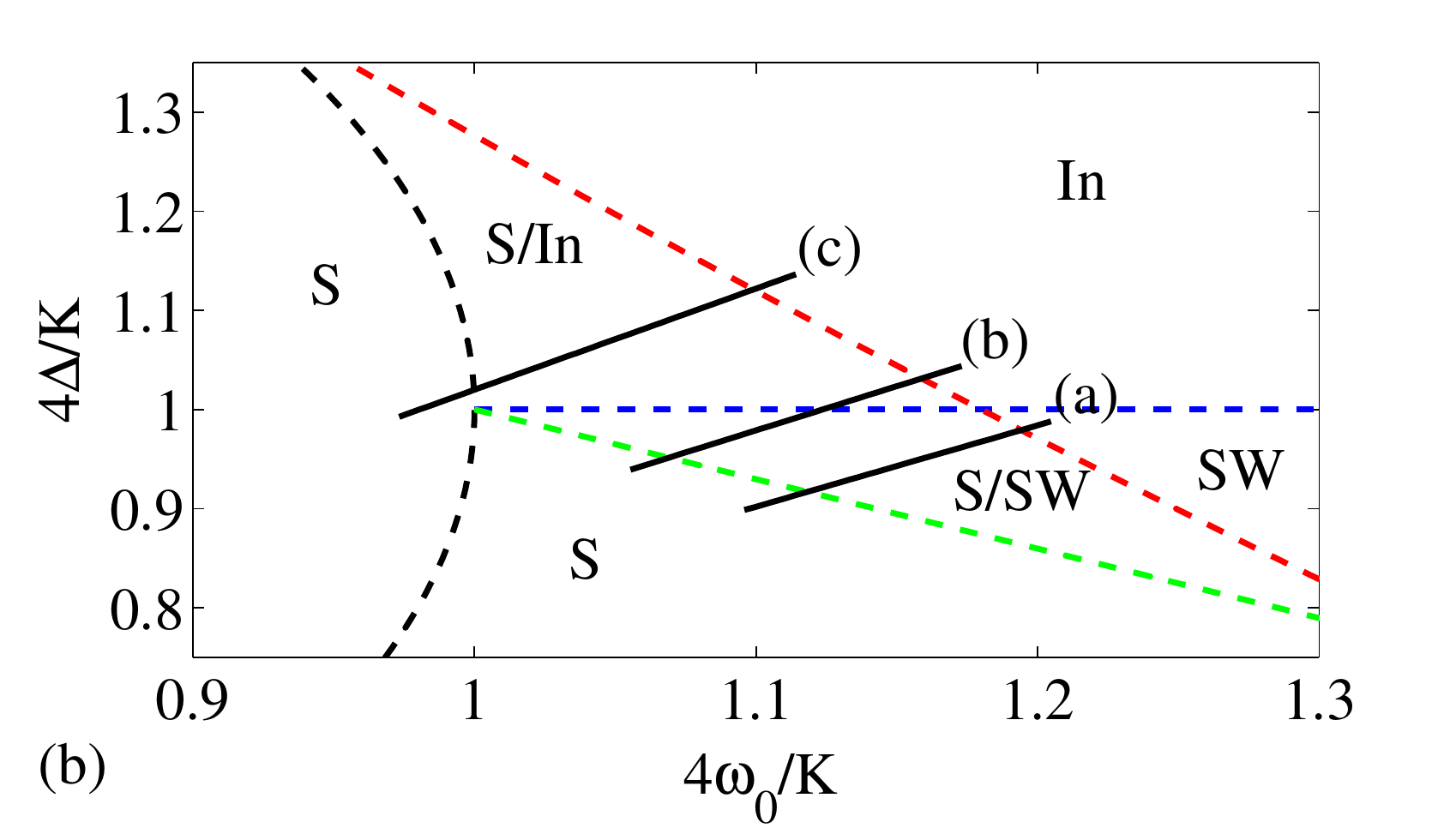}
\caption{(Color online) (a) Bifurcation diagram for the Kuramoto model with bimodal frequency distribution. Transcritical, Hopf, homoclinic, and saddle node/SNIPER bifurcations are plotted in dashed back, clue, green, and red, respectively. Paths taken in Fig.~\ref{bimodalexp} are ploted in solid black. (b) Zoomed-in view of bistable regions. We label regions where incoherent, synchronized, and standing-wave solutions are stable In, S, and SW, respectively.} \label{bimodalBD}
\end{figure*}

We simulate the system with $N_\sigma=2000$ oscillators in both communities with initial coupling strengths of $k_{11}=9$, $k_{22}=14$, and $k_{12}=k_{21}=0$ for values $\epsilon=0.105$ and $0.085$. In Fig. \ref{comms}(a) we plot $|r_{1}(t)|$ (solid blue curve) and $|r_2(t)|$ (dashed red curve) in the top panel and $k_{11}(t)$ (solid blue curve), $k_{22}(t)$ (dashed red curve), $k_{12}(t)$ (dot-dashed green curve), and $k_{21}(t)$ (dotted black curve) in the bottom panel for $\epsilon=0.105$. In Fig. \ref{comms}(b) we plot the same quantities for $\epsilon=0.085$. Although the two communities start in out-of-phase macroscopic states, for $\epsilon=0.105$ the macroscopic dynamics of the communities synchronize near $t=3T$. However, for $\epsilon=0.085$ they remain out of phase past $t=200T$. This significant difference in behavior for such a small change in $\epsilon$ suggests a sensitive dependence on the system parameters in addition to initial conditions.

\subsection{Bimodal frequency distribution}


Finally, we study uniform adaptation of a system of oscillators without time-delay, but having bistability due to a bimodal distribution of intrinsic frequencies.
The model we study is the following:
\begin{align}
\dot{\theta}_n&=\omega_n +k\text{Im}(re^{-i\theta_n}), \label{eqBimodalModel} \\
T\dot{k} &= G(k,z).\label{bi_k}
\end{align}
where $r=\frac{1}{N}\sum_{n=1}^Ne^{i\theta_n}$ is the normal Kuramoto order parameter and now we assume $\omega_n$ are drawn from the double Lorentzian 
\begin{equation}
g(\omega)=\frac{\Delta}{2\pi}\left[\frac{1}{(\omega-\omega_0)^2+\Delta^2}+\frac{1}{(\omega+\omega_0)^2+\Delta^2}\right],
\end{equation}
which is bimodal for $\Delta<\sqrt{3}\omega_0$. We note that in Ref.~\cite{So2} a similar oscillator system with bimodally-distributed frequencies is studied, but with an explicitly time-dependent sinusoidal coupling strength rather than system-dependent coupling adaptation. 

\begin{figure*}[t]
\centering
\includegraphics[width=0.31\linewidth]{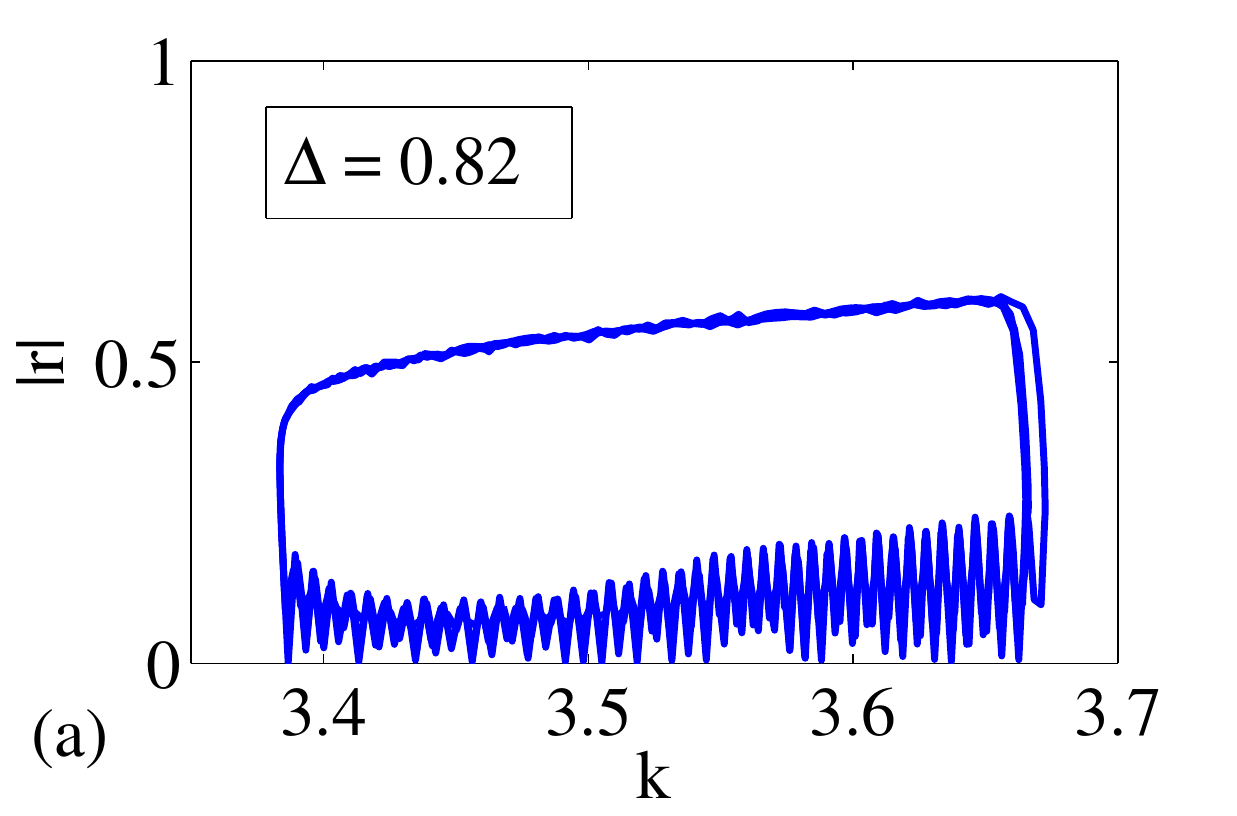} \hskip0.01\linewidth
\includegraphics[width=0.31\linewidth]{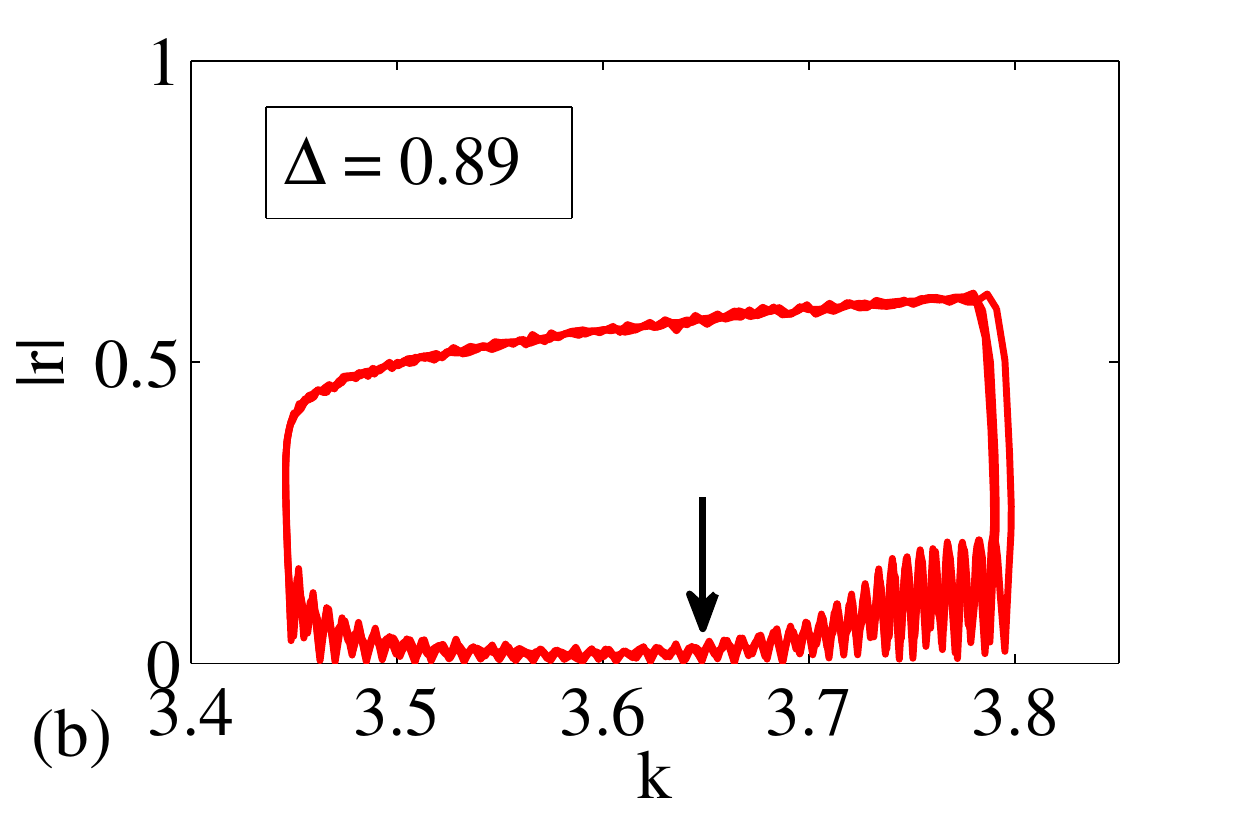} \hskip0.01\linewidth
\includegraphics[width=0.31\linewidth]{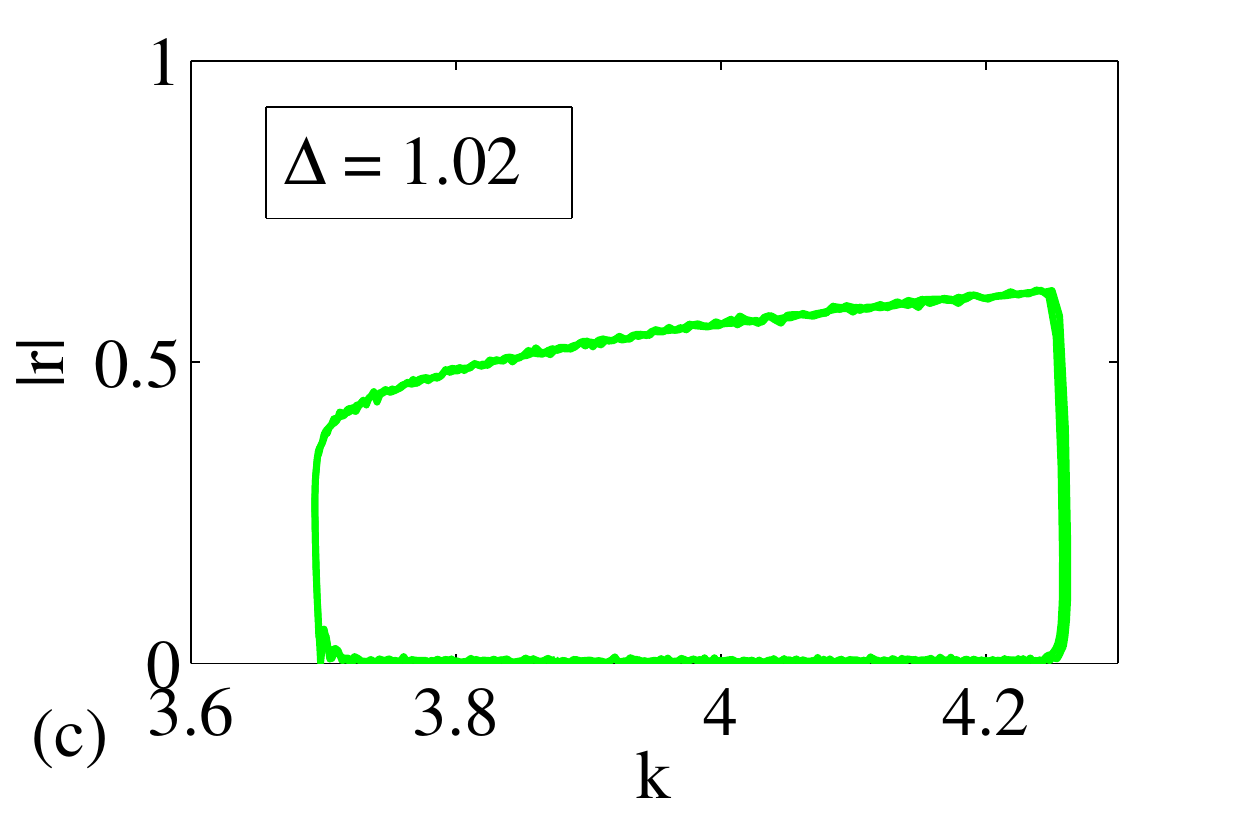}
\caption{(Color online) $|r(t)|$ vs $k(t)$ trajectories for the Kuramoto model with bimodal frequency distribution with uniform adaptation following Eq.~(\ref{eqkev}) for $N=2000$ oscillators, $\omega_0=1$, $\alpha=5$, $\beta=-5$, $\tau=1000$, and $\Delta=0.82$ (a), $0.89$ (b), and $1.02$ (c). A transition from incoherence to a standing wave solution in (b) is indicated by an arrow.} \label{bimodalexp}
\end{figure*}

This model is particularly interesting because in addition to the simple coherent and incoherent fixed points, stable solutions can also take the form of {\it standing waves} in which two synchronized groups [one corresponding to each peak of $g(\omega)$] oscillate with opposite angular velocity~\cite{Martens2}. These solutions are found for intermediate coupling strengths such that groups of oscillators with frequencies near $\omega_0$ and $-\omega_0$ synchronize, but these two groups do not synchronize with one another. These two groups act as giant oscillators that continue to pass one another, maximizing $|r|$ when the two groups have equal phase and minimizing $|r|$ when they have opposite phase. For a detailed analysis of this oscillator dynamics refer to Ref.~\cite{Martens2}.

In Fig.~\ref{bimodalBD}(a) we summarize the bifurcation diagram. Horizontal and vertical axes are $4\omega_0/k$ and $4\Delta/k$, respectively, and transcritical, Hopf, homoclinic, and saddle-node/SNIPER bifurcations are plotted in dashed black, blue, green, and red curves, and labelled TC, HB, HC, and SN/SNIPER, respectively. In Fig.~\ref{bimodalBD}(b) we show a zoomed-in view of the bistable regime and indicate regions where the incoherent, synchronized, and standing-wave solutions are stable. Regions are labelled S, In, and/or SW if the synchronized, incoherent, and/or standing wave solutions are stable in that region, respectively. For small $K$ the incoherent solution is the only stable solution. This solution loses stability either in a transcritical bifurcation or a Hopf bifurcation, giving rise to synchronized or standing-wave solutions. Synchronized solutions are also born at the saddle-node/SNIPER bifurcations, and the standing-wave solution disappears at the homoclinic bifurcation. There are two distinct regions of bistability in the approximately triangular area in the middle of the plot. For $4\Delta/k>1$ [labeled S/In in Fig.~\ref{bimodalBD}(b)] the synchronized and incoherent solutions are both stable, whereas for $4\Delta/k<1$ [labeled S/SW in Fig.~\ref{bimodalBD}(b)] the synchronized and standing-wave solutions are stable.

Letting Eq.~(\ref{bi_k}) take the linear form given by Eq.~(\ref{eqkev}) with $\alpha=5$, $\beta=-5$, and $\tau=1000$, we simulate a system with $N=2000$ oscillators for $\omega_0=1$ and (a) $\Delta=0.82$, (b) $0.89$, and (c) $1.02$. The respective trajectories in phase space (see solid black lines in Fig.~\ref{bimodalBD}) yield the following behaviors: (a) the system oscillates between synchronized and standing wave states; (b) the system repeats a synchronized$\to$incoherent$\to$standing wave $\to$ synchronized cycle; and (c) the system oscillates between synchronized and incoherent states. We plot the behavior of each in $(k,|r|)$ space in Figs. \ref{bimodalexp}(a), (b), and (c). Note in Fig.~\ref{bimodalexp}(b) that the macroscopic dynamics transition from incoherent to standing wave at $k\approx3.65$ (see arrow) as predicted by the Hopf bifurcation in Fig.~\ref{bimodalBD}. In this case we see three dynamical bifurcations in the transitions from incoherent$\to$standing-wave, standing-wave$\to$synchronized, and synchronized$\to$incoherent states.

\section{Discussion}

We have investigated analytically and numerically the effect of slow coupling adaptation on models of coupled phase oscillators exhibiting bistability and characterized complex macroscopic behavior that extends to other bistable phase oscillator systems where bistability arises (e.g., due to frequency adaptation~\cite{dane} or inertial terms~\cite{inertia}). In addition to states with simple macroscopic fixed points, we have observed for uniform coupling adaptation on bistable systems macroscopic excitable and intermittently synchronous states. We leave open the exploration of further dynamics that may occur for systems exhibiting multi-stability.


Besides considering only uniform coupling adaptation (i.e., allowing the global coupling strength of an all-to-all system to evolve depending on macroscopic system properties), we have also addressed network adaptation (i.e., allowing the links between individual oscillators to evolve according to their local properties). Network adaptation allows for heterogeneities in evolving networks to be accentuated and is often more realistic (e.g., Hebbian learning in neural systems \cite{Hebbian}). However, we have found that even when the underlying network structure is heterogeneous, which in turn promote heterogeneities in the coupling between oscillators, qualitatively similar macroscopic behavior emerges, i.e. fixed points, excitable, and intermittently synchronous states. Although our results for this case are purely numerical, we note that our results from the uniform adaptation model describe more heterogeneous networks with network adaptation very well. The development of more advanced methods for dimension reduction for heterogeneous oscillator networks is an open area of research, although progress continues \cite{ensemble}.

We also have considered uniform adaptation for systems with either community interaction or bimodal frequency distributions. In the community interaction model we have found complicated behavior even for simple parameter assumptions. We hypothesize that changing the manner in which communities interact and/or increasing the number of communities could lead to richer, more complicated dynamics, including chaotic macroscopic states. In the bimodal frequency distribution model, we have demonstrated new dynamic bifurcations corresponding to the transitions between standing-wave solutions and the typical incoherent and synchronized states.

This work also provides a strategy for reconciling the common disconnect between microscopic behavior (i.e. individual oscillator dynamics) and macroscopic phenomena. In the systems studied in this paper we have shown that entire populations of oscillators can combine into a single functional unit. For example, a wide range of parameters yields intermittent synchronous dynamics, which we liken to clock-like behavior. Similarly, we liken the dynamics of excitable and bistable states to neuron-like firing and switch-like behavior, respectively. One interesting direction of future research motivated by the work presented in this paper is the study of even more complex systems that are composed of many functional units in a hierarchical organization. In particular, one could study systems built out of different kinds of functional units, for instance to understand the resulting dynamics when networks of clocks, neurons, and switches interact. Because of their analytic tractability and simplicity, we believe that the results presented in this paper could prove a useful tool for understanding the generic behavior of these complex systems. 

\section*{Acknowledgements}
The authors would like to thank the Evolving Dynamical Systems organizing committee members for their invitation to participate in the special issue. Funding was provided in part by NSF Grant No. DMS-0908221.

\bibliographystyle{plain}


\end{document}